\documentclass[pre,twocolumn,showpacs,superscriptaddress,aps]{revtex4}

\usepackage{amsfonts}
\usepackage{amsmath}
\usepackage{amssymb}
\usepackage{graphicx}
\usepackage{dcolumn}
\usepackage{dsfont}
\usepackage{times}
\usepackage{color}

\begin{document}

\title{Acoustic solitons in waveguides with Helmholtz resonators: transmission line approach}  
 
\author{V. Achilleos}
\affiliation{Department of Physics, University of Athens, Panepistimiopolis,
Zografos, Athens 15784, Greece}
\author{O. Richoux}
\affiliation{LUNAM Universit{\'e}, Universit{\'e} du Maine, CNRS,
LAUM UMR 6613, Av. O. Messiaen, 72085 Le Mans, France}
\author{G. Theocharis}
\affiliation{LUNAM Universit{\'e}, Universit{\'e} du Maine, CNRS,
LAUM UMR 6613, Av. O. Messiaen, 72085 Le Mans, France}
\author{D. J. Frantzeskakis}
\affiliation{Department of Physics, University of Athens, Panepistimiopolis,
Zografos, Athens 15784, Greece}

\begin{abstract}

We report experimental results and study theoretically soliton formation and propagation in 
an air-filled acoustic waveguide side loaded with Helmholtz resonators. 
We propose a theoretical modelling of the system, which
relies on a transmission-line approach, leading
to a nonlinear  dynamical lattice model. The latter 
allows for an analytical description of the various soliton solutions for the pressure,
which are found 
by means of dynamical systems and multiscale expansion techniques. 
These solutions include Boussinesq-like and Korteweg-de Vries pulse-shaped 
solitons that are observed in the experiment, as well as nonlinear Schr\"{o}dinger 
envelope solitons, that are predicted theoretically. The analytical 
predictions are in excellent agreement with direct numerical simulations and in 
qualitative agreement with the experimental observations.

\end{abstract} 

 
\maketitle

\section{Introduction}

Solitons, namely robust localized waves propagating undistorted in nonlinear dispersive media 
\cite{Daux,rem,MJA}, have been studied extensively in various physical contexts. Indeed, soliton 
formation, stability, dynamics and interactions have been analyzed, both in theory and in 
experiments, in water waves \cite{johnson,infeld}, plasma physics \cite{infeld}, nonlinear 
optics \cite{kiag}, atomic Bose-Einstein condensation (BEC) \cite{bec}, and so on. 

On the other hand, solitons have also been studied in acoustics, both in solids and fluids 
\cite{book_acoustics}. In particular, nonlinear solitary waves have been the subject of many studies 
the last years in granular chains \cite{Nesterenko} and crystalline solids (see 
Ref.~\cite{Lomonosov} and references therein). In the latter case, solitons and solitary waves 
in crystals and their surfaces have been attained by nanosecond and picosecond laser ultrasonics 
methods. However, solitons in fluids have been studied less extensively: in fact, pertinent studies 
include seminal work by Sugimoto and co-workers, who studied theoretically \cite{sugi1,sugi2,sugi3} 
and demonstrated experimentally \cite{sugi2,sugi3} propagation of one-dimensional (1D) acoustic 
solitary waves in an air-filled waveguide, with a periodic array of Helmholtz resonators. 
In these works, the analysis was based on nonlinear wave equations 
with fractional derivative terms accounting for losses. For this model, soliton solutions 
were found in an implicit form, and turned out to be close to Korteweg-de Vries (KdV) solitons 
in some asymptotic limit; additionally, numerical studies on the model proposed in 
Refs.~\cite{sugi1,sugi2,sugi3} were recently reported too \cite{lomb}. Other relevant works 
include Refs.~\cite{jordan1}, where diffusive soliton solutions to the so-called 
Kuznetsov equation (which models weakly nonlinear acoustic wave propagation in 
viscoelastic media) were studied. 
Note that traveling wave solutions 
of a higher-order nonlinear acoustic wave equation of the Kuznetsov-type 
(valid for larger values of acoustic Mach number) were rigorously studied as well \cite{chen}. 
It is also relevant to mention the work of Ref.~\cite{noz}, where envelope solitons (holes) 
were predicted to occur in cylindrical acoustic waveguides (in this system, higher-order 
dispersive modes were taken into account). 


In this work, 
we revisit the theme of a lattice made of Helmholtz resonators side connected to a tube \cite{oliv}.
We present experimental observations of acoustic solitons in this setting, and 
propose an analytically tractable modeling, relying on an effective nonlinear 
transmission line (TL) description of the system. Our approach allows for both an efficient
description of the relevant experimental findings, and the prediction of other 
localized nonlinear structures that can be supported in the system.

Generally speaking, the TL approach is a powerful tool commonly used in electromagnetic (EM) wave 
applications \cite{davidson}, and has recently gained considerable attention due to its 
applicability in the analysis and design of both EM \cite{caloz} and acoustic \cite{bongard,fang,cheng} 
metamaterials. This approach also allows for the study of nonlinear effects, and particularly 
soliton formation and propagation, a theme that has been studied extensively in the past 
in the context of electrical TLs \cite{rem}, and more recently in the realm of 
TL metamaterials \cite{gv}. 


In our setting, namely the $1$D lattice of Helmholtz resonators, 
the proposed TL model correctly reproduces -- in the linear limit -- the dispersion relation. 
Furthermore, in the nonlinear regime, and in the small-amplitude, long-wavelentgth limit, 
the TL model describes -- in a good agreement with the experiment -- the 
soliton propagation in the waveguide. This simplified model also allows for  
an analytical study of the solitary waves based on universal nonlinear evolution equations 
that are derived by means of asymptotic expansions (see below). 
A direct numerical integration of the model provides numerical results that are 
consistent with the analytics and the experimental observations. Additionally, 
in the framework of the TL model, it is also possible to predict the 
formation of envelope solitons (both of the bright and the dark type).

We now proceed with a more specific description of our analysis and findings. 
First we note that our analysis relies on the study of an electrical 
TL, as per the electro-acoustic analogy, 
where the voltage corresponds to the acoustic pressure, and the current to the
volume velocity flowing through the waveguide's cross-sectional area \cite{book}. 
Nonlinear effects are taken into regard by incorporating nonlinear elements 
in the unit-cell circuit, accounting for the dependence of wave celerity on the pressure 
(note that Helmholtz resonators are assumed to have a linear response, while nonlinearity 
originates only for the large-amplitude wave propagation within the waveguide).  
This representation allows for the derivation of a nonlinear lattice model, which is 
studied numerically and analytically. In the numerical simulations, using initial conditions 
relevant to our experiments, we are able to reproduce soliton profiles and characteristics 
(speed, width, etc) in a good agreement with the experimental observations. Furthermore, 
employing the continuum approximation, we study analytically the lattice model, and show 
that it is intimately related (in proper temporal and spatial scales) to models 
that have been studied in the past in other branches of physics: these include 
a Boussinesq-type model and a KdV equation (originally used to describe shallow water waves  
\cite{johnson,MJA}, waves in plasmas \cite{infeld}, solitons in electrical TLs \cite{rem}, etc), 
as well as a nonlinear Schr\"{o}dinger (NLS) equation (describing deep water waves 
\cite{johnson,MJA}, optical solitons \cite{Daux,rem,MJA,kiag}, dynamics of BEC \cite{bec}, etc).
This way, we derive 
approximate pulse-like solitons of the Boussinesq and KdV type, as well as bright and dark 
envelope solitons satisfying an effective NLS equation. 
In all cases, we identify parameter regimes where different types of solitons can be formed, 
and present numerical results that are found to be in an excellent agreement with the 
analytical predictions. 

The paper is structured as follows. In Section~II, we describe the experimental setup and 
present experimental results for the formation of acoustic solitons in the $1$D lattice of Helmholtz 
resonators. We also introduce our model and, by employing the TL approach, 
derive the nonlinear lattice equation and compare numerical findings for the latter with 
relevant experimental results. Section~III is devoted to our analytical study: there, we present 
the various types of solitons that can be formed in our setting, identify 
relevant parameter regimes and spatio-temporal scales, and investigate their 
propagation characteristics. Finally, in Section~IV we present our conclusions and discuss 
future research directions.

\section{The Helmholtz resonator lattice}

\subsection{Experimental setup and observations}

\begin{figure}[tbp]
\includegraphics[width=8cm]{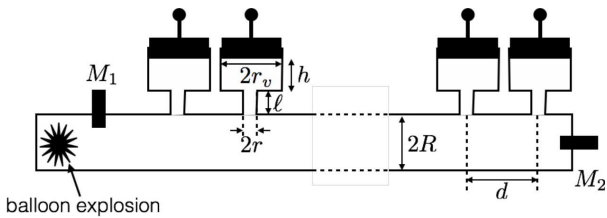}
\caption{\label{exper} Schematic illustration of the experimental setup.}
\end{figure}

We start by presenting our experimental setup, which consists of 
a long cylindrical waveguide, of length $L=6$~m, 
with a cross-section $S=\pi R^2$ with an inner radius $R=25\times10^{-3}$~m and a $5\times10^{-3}$~m thick wall. 
This waveguide is connected to an array of $60$ Helmholtz resonators, which are periodically
distributed. The distance between two consecutive resonators is $d=0.1$~m. Each resonator 
is composed by a neck (cylindrical tube with an inner radius $r=10\times10^{-3}$~m and 
a length $\ell = 20\times10^{-3}$~m) and a variable length cavity (cylindrical tube with 
an inner radius $r_v=21.5\times10^{-3}$~m and a maximum length $h=165\times10^{-3}$~m). 
Notice that the end of the waveguide,   
located at $d/2$ from the last resonator, is rigidly closed. 

The input signal is generated by the explosion of a balloon. The balloon is located at $20$~cm 
of the lattice into a waveguide connected to the main tube and is inflated until its explosion. 
The produced acoustic wave is measured with $2$ $PCB~106B$ microphones, carefully calibrated, 
which are located 
$20$~cm in front of the lattice and at the end of lattice (the microphone 
is embedded in the rigid end); recall that the propagation distance is $L=6.2$~m. 
The experimental setup is shown in Fig.~\ref{exper}.

\begin{figure}[tbp]
\includegraphics[width=8cm]{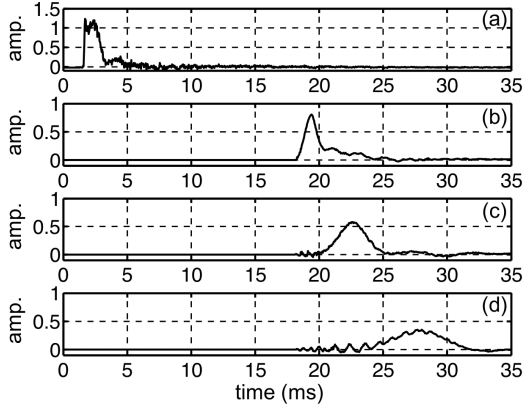}
\caption{
\label{exper2} 
Panel (a) shows the initial acoustic pressure, measured at $x=0$ m. 
Panels (b), (c) and (d) show, respectively, the acoustic pressure measured 
at the end of the lattice ($x=6.2$ m) for resonator cavity length $h=0.02$~m, $h=0.07$~m, 
and $h=0.165$~m.}
\end{figure}

Figure~\ref{exper2}(a) shows the temporal profiles of the normalized 
acoustic pressure measured at the first microphone located $20$~cm before the first resonator ($x=0$~m). 
The input signal, generated by the balloon explosion, can be described by a 
gate-signal with a large amplitude (around $30$~kPa) and a width around $1.5$~ms. 
Figures~\ref{exper2}(b),~\ref{exper2}(c) and~\ref{exper2}(d) present the temporal profiles 
of the acoustic pressure 
measured after a $6$~m propagation in the Helmholtz resonators lattice ($x=6.2$ m) 
for the cases of $h=0.02$~m, $h=0.07$~m and $h=0.165$~m respectively. Oppositely to the case of a 
waveguide without resonators where a shock wave is formed \cite{sugi3,Richoux_soliton}, we observe 
the propagation of a wave with a smooth shape through the lattice. The characteristics  
of this wave, namely shape, amplitude and velocity, are 
strongly dependent on the cavity length of the resonators, which defines the dispersion  
characteristics of the lattice (see Sec.~III.B). As it is seen, for $h=0.07$~m and $h=0.165$~m, 
the wave shape is clearly symmetrical, while for $h=0.02$~m this is not the case. 
Generally, it is observed that the competition between 
nonlinearities (due to a cumulative effect occurring for large amplitude pulse input) 
and dispersion in the medium (due to the presence of Helmholtz resonators) 
produces waves of constant shape, with amplitude 
dependent velocity, which are in fact acoustic solitons (note that we use the term ``soliton'' 
in a loose sense, without implying complete integrability \cite{MJA}).

\subsection{The discrete model: transmission line approach}

Next, in order to model our system and provide theoretical results for the above experimental 
observations of acoustic solitons, we will employ the TL approach. Our starting point relies 
on the consideration of an ideal fluid, and use of the fluid-dynamic equations, 
neglecting viscosity and other dissipative terms. If we restrict our analysis to 1D
flow -- as in the case of the experimental results of Fig.~\ref{exper2} -- wave propagation 
is described by the following equations:
\begin{eqnarray}
&&\frac{\partial \varrho}{\partial t} + \frac{\partial}{\partial x} (\varrho v) = 0,
\label{cont} \\ 
&&\frac{\partial v}{\partial t} + v \frac{\partial v}{\partial x} = -\frac{1}{\varrho}
\frac{\partial p}{\partial x}, \label{eul}
\end{eqnarray}
where $\varrho(p,s)$ is the fluid mass density, $s$ is the entropy, $v$ is 
the acoustic fluid velocity and $p$ is the acoustic pressure. 
We assume that the entropy $s$ is constant, while the mass 
density $\varrho$ and wave celerity
$c\equiv(\partial p / \partial \varrho)^{1/2}$ are considered as functions 
of the total pressure $p$. 
Accordingly, the acoustic fluid velocity $v$ can can be written as a single-valued function 
of the pressure $p$ so that 
$\partial v/\partial t = \left(d v/d p\right)\partial p/\partial t$.
We wish to model the acoustic propagation along the waveguide  
in the low frequency regime, where only plane waves can propagate, 
by means of the electro-acoustic analogy \cite{book}. 
Considering the long-wavelength limit, the mass conservation and Euler's equations 
(\ref{cont})-(\ref{eul}) between two points separated by $dx$ (much smaller than the acoustic 
wavelength) can be approximated as:
%
%
\begin{eqnarray}
u_n=C_w\frac{\partial p_{n+1}}{\partial t} +u_{n+1},
\label{mass2} \\
p_n=L_w \frac{\partial u_{n+1}}{\partial t} +p_{n+1},
\label{eul2}
\end{eqnarray}
where $u$ is the acoustic volume velocity and the subscripts $n$ and $n+1$ are related, 
respectively, to left and right side of the tube at some point $dx$.
According to the electrical analogy, the propagation along a unit-cell with length $dx$ can be 
modelled by a simple electrical circuit for the ``current'' $u_n$ and the ``voltage'' $p_n$, 
consisting of an inductance $L_w$ and a capacitance $C_w$. In the linear regime, these are given by:
\begin{eqnarray} 
L_{w0}=\varrho_0 dx/S,\quad C_{w0}=S dx/\varrho_0 c_0^2,
\label{Cap}
\end{eqnarray}
where $\varrho_0$ is the density evaluated at the equilibrium state, 
$c_{0}$ is the speed of sound.  
Notice that in the nonlinear regime $L_w$ and $C_w$ can define a wave celerity as 
$c_{\rm NL}^2 = 1/L_w C_w$. 
For our analysis below, we will assume that 
the inductance is linear, $L_w=L_{w0}$, while the capacitance defined as $C_w=S dx/\varrho_0 c_{\rm NL}^2$ is nonlinear, 
depending on the pressure $p$; 
this choice, relies on the approximation that (to a first order) 
the density does not depend on $p$, while the wave celerity $c_{\rm NL}$ depends 
on $p$. 

\begin{figure}[tbp]
\centering
\includegraphics[width=6cm]{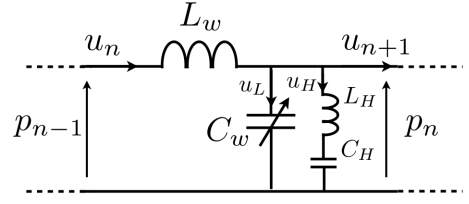}
\caption{\label{fig_cell_lattice} The unit-cell circuit of the nonlinear Helmholtz lattice model.}
\end{figure}

In order to model the experimental setup that incorporates the Helmholtz resonators, 
we will include an additional parallel branch in the unit-cell circuit, composed by a serial 
combination of an inductance $L_H$ and a capacitance $C_H$, as shown in Fig.~\ref{fig_cell_lattice}. 
We consider the response of the Helmholtz resonators to be linear. Nonlinearity originates 
only from the large amplitude acoustic propagation within the waveguide. Thus, 
in the low frequency approximation, the relevant 
inductance and capacitance are given by $L_H=\varrho_0 \ell/S_n$ and
$C_H=V_H/\varrho_0 c_{0}^2$, respectively, where $\ell$, $S_n$ and $V_H$ are the length
and the cross-sectional area of the resonator neck, and the total volume of the resonator 
cavity, respectively. Notice that, by including a resonator in each unit-cell, it 
is natural to set $dx = d$ (recall that $d$ is the distance between two successive resonators). 

Using the unit-cell circuit of Fig.~\ref{fig_cell_lattice}, we can now use Kirchhoff's voltage and 
current laws and derive an evolution equation for the pressure $p_n$ in the $n$-th cell 
of the lattice. Let us first consider the Kirchhoff's voltage law for two successive cells, 
which yields:
\begin{eqnarray}
p_{n-1}-p_n&=&L_w\frac{d}{dt}u_n, 
\label{eq1_V} \\
p_{n}-p_{n+1}&=&L_w\frac{d}{dt}u_{n+1}.
\label{eq2_V}
\end{eqnarray} 
Subtracting the above equations, we obtain the difference equation:
\begin{eqnarray}
\hat{\delta}^2p_{n}&=&\hat{\mathcal{L}}(u_n-u_{n+1}), 
\label{eq3_V}
\end{eqnarray} 
where $\hat{\delta}^2p_{n}\equiv p_{n+1}-2p_n+p_{n-1}$ and $\hat{\mathcal{L}}\equiv L_w d/dt$.
On the other hand, Kirchhoff's current law yields:
\begin{equation}
u_{n}-u_{n+1}=\frac{d}{dt}(C_w p_n) + \hat{P}^{-1}\frac{d p_n}{dt},
\label{eq2_I}
\end{equation}
where the first and second terms in the right-hand side denote the currents across 
the capacitance $C_w$ and the Helmholtz branch, respectively, with 
$\hat{P}^{-1}$ being the inverse of the operator 
$\hat{P} \equiv L_H d^2/dt^2 + 1/C_H$.

Substituting Eq.~(\ref{eq2_I}) into Eq.~(\ref{eq3_V}), we obtain the following equation for the 
pressure $p_n$: 
\begin{align}
&L_w C_H\frac{d^2 p_n}{dt^2}-\left(1+L_HC_H\frac{d^2}{dt^2}\right)\hat{\delta}^2 p_n \nonumber
\\
&+L_w\frac{d^2}{dt^2}\left(1+L_HC_H\frac{d^2}{dt^2}\right)(C_wp_n)=0,
\label{eq22_final}
\end{align}
where it is reminded that the capacitance $C_w$ depends on the pressure. In order to quantify this 
dependence, 
%
%
and take into account the nonlinear 
processes in the propagation, we can add a nonlinear term in the celerity as \cite{book,Ham}:
\begin{eqnarray}
c_{\rm NL} \approx c_0(1+\beta_0 p /\varrho_0 c_0^2),
\label{celer}
\end{eqnarray}
where $c_0=343.26$~m/s is the speed of sound at room temperature, 
and $\beta_0=1.2$ 
for the case of air. Then, 
the second of Eqs.~(\ref{Cap}) leads to the following pressure-dependent capacitance $C_w$:
\begin{equation}
C_w(p_{n})\approx C_{w0}+C'_w p_{n},
\label{nlinear_capacitance}
\end{equation} 
where $C_{w0}=Sd/\varrho_0 c_0^2$ is a constant capacitance (relevant to the linear case) and 
$C'_w=-2\frac{\beta_0}{\varrho_0 c_0^2}C_{w0}$. 
Substituting Eq.~(\ref{nlinear_capacitance}) into Eq.~(\ref{eq22_final}), we obtain the equation:
\begin{eqnarray}
\!\!\!\!\!\!\!\!\!\!\!
&&\frac{d^2 p_n}{d t^2}-\frac{c_0^2}{\kappa d^2}
\left(1+\frac{1}{\omega_0^2}\frac{d^2}{dt^2}\right)\hat{\delta}^2 p_n 
\nonumber \\
\!\!\!\!\!\!\!\!\!\!\!
&&+\frac{1}{\kappa}\frac{d^2}{dt^2}\left(1+\frac{1}{\omega_0^2}\frac{d^2}{dt^2}\right)
\left[p_n\left(1-2\frac{\beta_0}{\varrho_0 c_0^2} p_n\right)\right]=0,
\label{dimens1}
\end{eqnarray}
where $\omega_0=c_0\sqrt{S_n/S_v h \ell}$ is the Helmholtz resonance frequency 
($S_n=\pi r^2$ and $S_v=\pi r_v^2$ are the cross-sectional areas of the resonator neck and cavity,
respectively), $\kappa=V_H/V$ is a geometrical factor 
(ratio of the volume of the Helmholtz resonator $V_H$ over the 
tube volume $V$ in a unit cell of length $d$, and we have used 
the following equations connecting the transmission line parameters 
with the acoustic waveguide characteristics:
\begin{align}
L_wC_H=\frac{\kappa}{d^2c_0^2}, &\quad L_HC_H=\frac{1}{\omega_0^2}, 
& L_wC_{w0}=\frac{d^2}{c_0^2}, & 
\end{align}
where $L_w$ is also evaluated at $\varrho=\varrho_0$.
The above nonlinear dynamical lattice equation is one of the main results of the present work: 
it describes the propagation of acoustic waves in a tube with an array of Helmholtz resonators. 
This simplified model will be used below in order to derive analytical solitary wave solutions 
that are supported in this setting --as is also evident from the experimental results shown in Fig.~\ref{exper2}. 

\subsection{Comparison with the experiment}

Let us now proceed by comparing results that can be derived 
in the framework of the lattice model~(\ref{dimens1}) with the experimental results presented above.

We numerically integrate Eq.~(\ref{dimens1}) by means of a 4th-order Runge-Kutta method, 
using an initial condition similar to the experiments, as shown in the top panel of
Fig.~\ref{exper2}. In particular, we use a super-Gaussian pulse of the form 
\begin{equation}
p_{n=0}(t)=A\exp\left[-\left((t-t_0)/w_0\right)^{16}\right],
\end{equation}
of amplitude $A=30$~kPa and width $w_0/2=400$~Hz. The values of the coefficients of the 
various terms of Eq.~(\ref{dimens1}) depend actually only on the cavity length $h$, since all 
other parameters are fixed. 

The results of the direct numerical simulations, corresponding to the three different cavity 
lengths used in the experiment ($h=0.02,~0.07,~0.165$~m), are shown in Fig.~\ref{expernumer}. 
In all cases shown in panels (b)-(d) of Fig.~\ref{expernumer}, the profile of each pulse is 
shown for the lattice cite $n=60$, corresponding to a distance $6$~m from the first resonator.  
The profiles in panels (b)-(d) are time shifted by $\Delta t\approx 0.9$~ms corresponding to 
the propagating time needed for the initial pulse to reach the first resonator; this is done
to facilitate direct comparison with the experimental results of Fig.~\ref{exper2}.

Comparing corresponding panels of Figs.~\ref{exper2} and \ref{expernumer} for each of the three 
different values of $h$, it is seen that the solitary waves 
obtained numerically via Eq.~(\ref{dimens1}) have approximately the same width as those observed 
in the experiment. Notice that quantitative differences between numerical and experimental soliton 
amplitudes, as well as the presence of ``tails'' attached to the solitons 
(which are absent in the experimental data), may be qualitatively understood by 
(i) the presence of losses in the experiment [which are not included in the simplified model of Eq.~(\ref{dimens1})],
and (ii) the fact that the initial conditions used in the experiment and simulations are different.

In any case, the above comparison shows that Eq.~(\ref{dimens1}) can be used to 
describe, in a fairly good agreement with the experiment, the formation 
of acoustic solitary waves. Below we will show that, using this simplified model, we can obtain  
analytically different types of acoustic solitons in different experimentally relevant regimes. 

\section{Acoustic solitons}

\subsection{The continuum approximation}

For our analytical considerations, we will focus on the continuum limit 
of Eq.~(\ref{dimens1}), corresponding to $n\rightarrow\infty$ and $d \rightarrow 0$ 
(but with $nd$ being finite); in such a case, the pressure becomes 
$p_n(t) \rightarrow p(x,t)$, where $x=nd$ is a continuous variable. 
Then, the difference operator $\hat{\delta}^2$ is approximated by 
$\hat{\delta}^2p_n \approx d^2p_{xx}$, where terms of the order $\mathcal{O}(d^4)$ and higher 
are neglected, and subscripts denote partial derivatives. It is also convenient 
to express our model in 
dimensionless form; this can be done upon introducing the normalized variables 
$\chi$ and $\tau$ and normalized pressure $P$ 
[of order $\mathcal{O}(1)$], which are defined as follows: 
\begin{equation}
\tau=\tilde{\omega}_0 t, \qquad
\chi=\frac{\tilde{\omega}_0}{c_0\sqrt{\alpha}}x, \qquad 
\frac{p}{p_0} = 
\epsilon P,
\end{equation}
where $\tilde{\omega}_0$ is a characteristic spectral width or inverse temporal width 
(which is set by the  initial condition), 
$p_0=\varrho_0 c_0^2/2\beta_0$, $\alpha=1/(1+\kappa)$, and $\epsilon$ is a 
dimensionless small parameter ($\epsilon \ll 1$), defining the strength of the nonlinearity. 
In these variables, the continuum limit of Eq.~(\ref{dimens1}) reads:
\begin{figure}[tbp]
\includegraphics[trim= 0 2.8in 0 3in,width=8cm]{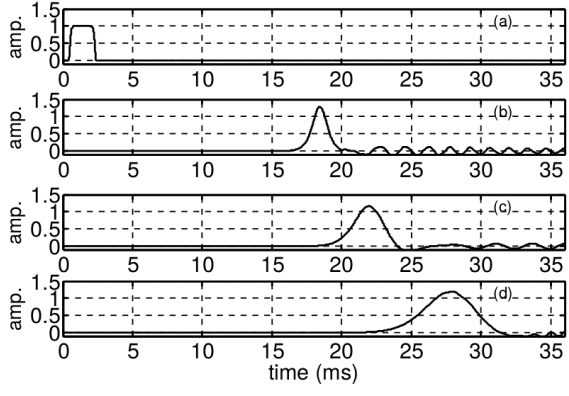}
\caption{
\label{expernumer} 
Top panel: Initial condition $p_0$ used for the numerical integration of Eq.~(\ref{dimens1}).
Rest of the panels (time shifted by $0.9$~ms --see text) show the pressure at $n=60$ 
as a function of time, for the same value of the cavity length
as in the respective experimental data (cf. Fig.~\ref{exper2}).}
\end{figure}
\begin{eqnarray}
P_{\tau\tau}&-&P_{\chi\chi}-\Omega^2(P_{\chi\chi\tau\tau}-\alpha P_{\tau\tau\tau\tau}) \nonumber \\
&-&\epsilon\alpha\left[(P^2)_{\tau\tau}+\Omega^2(P^2)_{\tau\tau\tau\tau}\right]=0,
\label{eq_final_d}
\end{eqnarray}
where $\Omega=\tilde{\omega}_0/\omega_0$. 
Equation~(\ref{eq_final_d}) is a 
Boussinesq-like model, which has been originally proposed for studies of solitons 
in shallow water \cite{johnson,MJA}, but later was used in studies of solitons in different 
contexts, including electrical TLs \cite{rem}. In our case, 
the dispersion terms of Eq.~(\ref{eq_final_d}) are due to 
the presence of Helmholtz resonators, 
and their strength is measured by the dimensionless parameter $\Omega$. 
The strength of the nonlinear terms, on the other hand, is set by the parameter $\epsilon$. 
Notice that in the absence of the Helmholtz resonators, i.e., for $\omega_0 \rightarrow \infty$ 
and $\kappa=0$ (i.e., $\Omega=0$ and $\alpha=1$), Eq.~(\ref{eq_final_d}) is reduced to 
the well-known Westervelt equation, which is a common nonlinear model describing 1D acoustic wave 
propagation \cite{Ham}.

\subsection{Linear theory}

We start by considering the linear limit of Eq.~(\ref{eq_final_d}) and the 
respective dispersion relation. Note that in the limit of $\epsilon \rightarrow 0$, 
Eq.~(\ref{eq_final_d}) is reduced to the linear wave equation (in the lossless case) 
studied in Ref.~\cite{sugilin} (see Eq.~(61) of this work).

\begin{figure}[tbp]
\includegraphics[width=8.5cm]{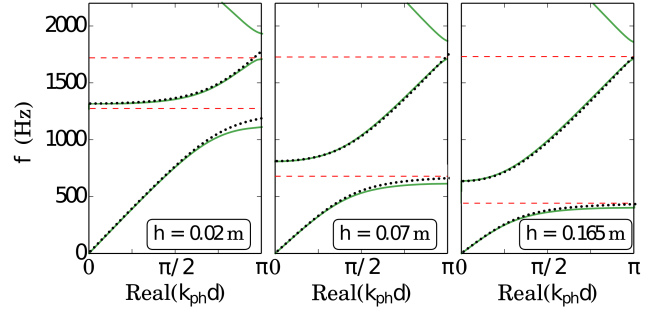}
\caption{\label{dispfig} (Color online) 
The dispersion relation, expressed in physical units, as obtained via Eq.~(\ref{eq2})
[solid (green) line], for three different values of the Helmholtz resonator cavity length $h$. 
This result is compared to the approximate one of Eq.~(\ref{dispers3}) [dotted (black) line]. 
The lower and upper horizontal dashed (red) lines depict the Helmholtz resonance frequency 
$f_0 \equiv \omega_0/2\pi$, and the Bragg frequency $f_B$, respectively. Note that $f_0$ takes the values 
$1270$~Hz ($h=0.02$~m), $679$~Hz ($h=0.07$), and $442$~Hz ($h=0.165$~m), while in all cases $f_B$ 
is fixed from the lattice constant $d=0.1$~m, and takes the value $f_B=c_0/2d = 1720$~Hz.}
\end{figure}

Assuming propagation of plane waves in the lattice, 
of the form $P \propto \exp[i(k\chi-\omega \tau)]$, we obtain the following 
dispersion relation connecting the wavenumber $k$ and frequency $\omega$:
\begin{eqnarray}
D(\omega,k)\equiv k^2-\omega^2-\Omega^2(k^2\omega^2-\alpha\omega^4)=0.
\label{dispers2}
\end{eqnarray}
Since all quantities in the above dispersion relation are dimensionless, it is also relevant to 
express Eq.~(\ref{dispers2}) in physical units. In particular, taking into regard that the 
frequency $\omega_{\rm ph}$ and wavenumber $k_{\rm ph}$ in physical units are connected with 
their dimensionless counterparts through $\omega=\omega_{\rm ph}/\tilde{\omega}_0$ and $k=k_{\rm ph}c_0\sqrt{\alpha}/\tilde{\omega}_0$, we can express Eq.~(\ref{dispers2}) in 
the following form:
\begin{eqnarray}
k_{\rm ph}^2-\frac{\omega_{\rm ph}^2}{c_0^2\alpha}-\frac{1}{\omega_0^2}
\left(k_{\rm ph}^2\omega_{\rm ph}^2-\frac{\omega_{\rm ph}^4}{c_0^2}\right)=0.
\label{dispers3}
\end{eqnarray}
%
%
%
Solving Eq.~(\ref{dispers3}) analytically with respect to $k_{\rm ph}$, we can then determine 
the frequency $f=\omega_{\rm ph}/2\pi$ as a function of the normalized wavenumber $k_{\rm ph}d$, 
and plot the resulting dispersion relation. 
The relevant result is depicted in Fig.~\ref{dispfig} by the dotted (black) line, 
for the three different values of the Helmholtz resonator cavity length $h$ used in the 
experiment, namely $h=0.02$~m, $h=0.07$~m and $h=0.165$~m.

On the other hand, the solid (green) line in the same figure shows the respective result 
(for the lossless case under consideration) for the dispersion relation, as 
obtained using Bloch theory and the transfer matrix method \cite{sugilin}:
\begin{equation}
\label{eq2}
  \cos (k_{\rm ph}d) = \cos\left(\frac{\omega_{\rm ph}}{ c_{0}}d\right)
  + i\frac{Z_{0}}{2Z_{b}}\sin\left(\frac{\omega_{\rm ph}}{ c_{0}}d\right),
\end{equation}
where $Z_{b}$ is the input impedance of the Helmholtz resonator branch, and 
$Z_{0}=\varrho_0 c_{0}/S$ the acoustic characteristic impedance of the waveguide;  
for the lossless case $Z_{b}=i(\omega_{\rm ph} L_H-1/\omega_{\rm ph} C_H)$. Note that,  
in the linear regime, the transmission line approach to acoustic waveguides with periodically arranged 
Helmholtz resonators, has also been proposed and discussed in other works (see, e.g., Refs.~\cite{fang,cheng}).

The dispersion relation (\ref{eq2}) obviously reflects the periodicity of the system, featuring 
a band-gap structure. This becomes clear upon observing the upper gap shown in Fig.~\ref{dispfig}, which 
originates from Bragg-type constructive interference of reflections, and is characterized by the 
Bragg frequency $f_B=c_0/2d$; the latter is equal to $1720$~Hz for our setting, and is depicted by the upper 
dashed (red) lines in the three panels of Fig.~\ref{dispfig}. In addition, the dispersion relation features 
still another gap (usually called ``resonator'' or ``hybridization'' band gap), 
originating from Fano resonances/interference, due to the presence of the Helmholtz resonators.
This gap is around the resonance frequency of the Helmholtz resonator, $f_0=\omega_0/2\pi$, which is 
chosen to be sufficiently smaller than the Bragg frequency $f_B$; such a choice is possible by properly 
fixing the cavity length $h$. The location of $f_0$ for the three different values of $h$ that are used 
in the experiment is depicted by the lower dashed (red) lines in the three panels of Fig.~\ref{dispfig}. 
Observing the structure of the first (lower) band, it is clear that increasing the Helmholtz cavity length, 
the resonance frequency decreases, and additionally the dispersion in the low-frequency regime increases. 
On the other hand, observing the structure of the second band, it is evident that the increase of the 
Helmholtz cavity length results in a decrease of dispersion near the Brillouin boundary.

Comparing the dispersion relation (\ref{eq2}) with the one resulting from the 
continuum approximation [cf. Eq.~(\ref{dispers3})], we find a very good agreement between the two, 
especially in the regime of low frequencies (note that in this regime the transmission line approach is expected 
to be more accurate). In particular, the dispersion relation (\ref{dispers3}) 
is able to follow the lower band, the first gap and the second band, especially in the regime of $k_{\rm ph}d \ll 1$ 
(where the continuum approximation is formally more accurate). For instance, the 
upper band gap edge for $k_{\rm ph}=0$ is found from Eq.~(\ref{dispers3}) as  
$\omega_{\rm ph} = \omega_0/\sqrt{\alpha}$, which is in agreement with the effective medium approach of 
Ref.~\cite{fang}. 

In addition, the result of the continuum approximation is still in reasonable agreement 
with the result of Eq.~(\ref{eq2}) for moderate and larger values of $k_{\rm ph}d$, even sufficiently close to the 
Brillouin boundary. For the lower band, this agreement can be attributed to the fact that the first gap 
is only due to the Helmholtz resonance and not due to the system's periodicity. In other words, dispersion 
only comes into play due to Helmholtz resonance (recall that dispersion in Eq.~(\ref{eq_final_d}) 
vanishes for $\Omega=0$ or $\omega_0 \rightarrow \infty$). As concerns the 
second band, it can be observed that, for sufficiently large $k_{\rm ph}$, the dispersion relation (\ref{dispers3}) 
becomes $\omega_{\rm ph}=k_{\rm ph}c_0$, for every cavity length $h$. The same behavior is also found from Eq.~(\ref{eq2}), 
which can explain the agreement with Eq.~(\ref{dispers3}), even close to the 
Brillouin boundary (at least for the parameter values used in the experiment). 
There, it is obvious that the continuum approximation becomes invalid, because 
the dispersion relation (\ref{dispers3}) does not take into regard the periodicity of the system, 
thus failing to capture the band gap around $f_B$ (as well as the structure of the spectrum for $f>f_B$). 
Notice that this failure is more pronounced 
for smaller values of cavity length (cf. left panel of Fig.~\ref{dispfig}) due to the fact that, in this case,  
periodicity-induced dispersion is enhanced.

Thus, concluding this section, the continuum approximation Eq.~(\ref{eq_final_d}) is quite accurate 
in capturing the (Helmholtz resonance-induced) dispersion properties of the system in the low-frequency and 
long-wavelength regimes -- as is the case for the parameter values used in the experiment. 
It is thus reasonable to expect that different types of solitons may be obtained in different 
regimes of the dispersion relation, by exploiting the relative strength between dispersion and 
nonlinearity. A relevant study is appended in the following sections.

\subsection{Boussinesq and KdV pulse-like solitons}

First we focus on the regime where the dispersion and nonlinearity terms of Eq.~(\ref{eq_final_d}) 
are of the same order, i.e., $\epsilon \sim \Omega^2$. Given that we have already assumed 
a weak nonlinearity, it is obvious that the last term in the left-hand side of 
Eq.~(\ref{eq_final_d}), which is $\propto \epsilon\Omega^2$ can be neglected. 
In such a case, Eq.~(\ref{eq_final_d}) is reduced to the following equation:
\begin{eqnarray}
\!\!\!\!\!\!\!\!\!  
P_{\tau\tau}-P_{\chi\chi}-\Omega^2(P_{\chi\chi \tau\tau}-\alpha P_{\tau\tau\tau\tau})-\epsilon\alpha(P^2)_{\tau\tau}=0, 
\label{bouss}
\end{eqnarray}
which is actually a combination of the so-called bad and improved Boussinesq equation 
(see, e.g., Ref.~\cite{rosenau} for the definition and discussion of these models). 
Travelling wave solutions of the above equation 
can readily be obtained by introducing the ansatz $P(\chi,\tau)=\Phi(\xi)$, 
where $\xi=\delta(\tau-\chi/v)$, while $v$ and $\delta$ denote 
the velocity and inverse width of the wave. 
Then, assuming vanishing boundary conditions for $\Phi$, namely $\Phi\rightarrow 0$ 
as $|\xi|\rightarrow \infty$, we derive from Eq.~(\ref{bouss}) the following ordinary 
differential equation (ODE) for $\Phi(\xi)$:
\begin{eqnarray}
A\Phi''+B\Phi-\epsilon\alpha \Phi^2=0,
\label{mkdv2}
\end{eqnarray}
where primes denote differentiation with respect to $\xi$, while 
$A=\Omega^2\left(\alpha -1/v^2\right)$ and $B=1-1/v^2$. Equation~(\ref{mkdv2}) 
can be seen as an equation of motion of a particle in the presence of the potential 
$V(\Phi)=(B/2A)\Phi^2-(\epsilon\alpha/3A)\Phi^3$. A straightforward analysis shows
that the only physically relevant solution, with the correct (vanishing) boundary conditions, 
corresponds to a homoclinic orbit, for $A<0$, $B>0$, relevant to the hyperbolic fixed point 
$\Phi=3B/2\epsilon \alpha$. This solution reads:	
\begin{eqnarray}
\!\!\!\!\!\!\!\!\!\!\!
P(\chi,\tau)&=&\left(\frac{\Omega^2}{\epsilon}\right)\left(\frac{6\kappa\delta^2}{1+4\delta^2\Omega^2}\right)
{\rm sech}^2\left[\delta\left(\tau-\frac{\chi}{v}\right)\right],
\label{exactsoliton}
\end{eqnarray}
where the velocity is given by $v=[(1+4\delta^2\Omega^2)/(1+4\alpha\delta^2\Omega^2)]^{1/2}$. 
Obviously, the above solution is characterized by one free parameter, the inverse width $\delta$. 
Note that since $\Omega^2/\epsilon\sim 1$ (as per our assumption above), the free parameter 
$\delta$ is also $\sim 1$. Thus, the normalized pressure $P$, along with its spectral width, 
are of the order of unity as well. 
Using Eq.~(\ref{exactsoliton}), we can express --for the sake of clarity-- the corresponding 
approximate solution of Eq.~(\ref{dimens1}) in terms of the original space and time coordinates 
as follows:
\begin{eqnarray}
\!\!\!\!\!\!\!\!
\frac{p(x,t)}{p_0}&\approx &\frac{3\kappa\delta^2(\tilde{\omega}_0/\omega_0)^2}{1+4\delta^2(\tilde{\omega}_0/\omega_0)^2}~{\rm sech}^2
\left[\delta\tilde{\omega}_0\left(t-\frac{x}{v}\right)\right]. 
\label{exactphys}
\end{eqnarray}
Notice that, in physical units, the velocity of the soliton reads:
\begin{equation}
v=c_0\sqrt{\alpha}\sqrt{\frac{\omega_0^2+4\delta^2\tilde{\omega}_0^2}
{\omega_0^2+4\alpha\delta^2\tilde{\omega}_0^2}},
\label{bousvel}
\end{equation}
and is bounded (as follows from the requirements $A<0$ and $B>0$ mentioned above) 
according to:
\begin{equation}
c_0\sqrt{\alpha} < v < c_0.
\end{equation}
This shows that the velocity of the Boussinesq-type soliton of Eq.~(\ref{exactphys}) is 
lower than the speed of sound (i.e., the soliton is subsonic), in accordance with the 
analysis of Ref.~\cite{sugi3} for small geometrical factor $\kappa$ [see Eq.~(2.14) of this work].

\begin{figure}[tbp]
\includegraphics[width=8cm]{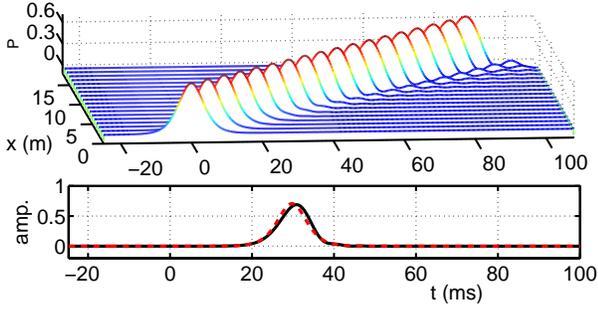}
\caption{\label{exact1} Top panel: 3D plot depicting the evolution of a soliton 
of the form of Eq.~(\ref{exactsoliton}), obtained by numerically integrating 
Eq.~(\ref{dimens1}) for a distance corresponding to $200$ sites (physical distance 
$x=20$~m. 
The bottom panel shows the temporal profile of the normalized pressure, $p/p_0$, at the site 
$n=60$. Parameter values correspond to the 
experimental ones, for a Helmholtz resonator with cavity length $h=0.07$ m. 
The dashed (red) line in the bottom panel depicts the analytical 
result of Eq.~(\ref{exactsoliton}), while the solid (black) line the result of the simulation.}
\end{figure}

We have numerically integrated the nonlinear lattice model of Eq.~(\ref{dimens1}), 
using as an initial condition, $p_1$ (i.e., the pressure at the first site of the lattice),
the functional form of the soliton of Eq.~(\ref{exactphys}) at $x=0$; we have used the parameter 
values $\delta\tilde{\omega}_0=0.1$, and a cavity length $h=0.07$~m. The results of our 
simulations are shown in Fig.~\ref{exact1}. The top panel shows a three-dimensional (3D) plot 
depicting the evolution of the pressure $p$, while the bottom panel shows the temporal 
profile of the pressure at the lattice site $n=60$, corresponding to a physical distance 
$x=6$~m. It is observed that the soliton propagates for about $20$~m with almost no distortion. 
In fact, the only noticeable effect is a small amount of radiation emitted by the soliton 
during its evolution (cf. the structure formed at the leading edge of the pulse); this 
effect can naturally be attributed to the fact that Eq.~(\ref{exactphys}) 
is nothing but an approximate solution --derived in the continuum limit-- of the 
lattice model of Eq.~(\ref{dimens1}). Nevertheless, as is also shown in the bottom panel of 
Fig.~\ref{exact1}, our analytical approximation is very good --at least for propagation 
distances up to $40$~m: indeed, the analytical result [dashed (red) line] for the soliton 
profile (at $x=6$~m) in the bottom panel of the figure, 
almost coincides with the corresponding numerical result [solid (black) line]. 

For longer propagation distances ($x\gtrsim 40$~m), however, the continuous emission 
of radiation of the Boussinesq-type solitons eventually lead to their disintegration. 
More robust soliton solutions --in the 
same parametric region-- can be obtained upon considering the long-wavelength, far-field limit 
of the Boussinesq-type Eq.~(\ref{eq_final_d}), which is the KdV equation. Indeed, 
using a formal multiscale expansion method, we can reduce Eq.~(\ref{eq_final_d}) 
to a KdV equation, and use the latter to derive approximate solutions of Eq.~(\ref{dimens1}). 
We thus proceed upon using the slow variables:
\begin{eqnarray}
T=\epsilon^{1/2}(\tau-\chi),\qquad 
X=\epsilon^{3/2}\chi,
\label{slow}
\end{eqnarray}
and express Eq.~(\ref{eq_final_d}) as follows:
\begin{eqnarray}
&&2\epsilon^2P_{XT}-\epsilon^3P_{XX}-\Omega^2\big(\epsilon^2P_{TTTT}
-2\epsilon^3P_{XX}+\epsilon^4P_{XX} \nonumber \\
&&-\alpha\epsilon^2P_{TTTT}\big)
-\epsilon^2\alpha\left[(P^2)_{TT}+\epsilon(P^2)_{TTTT}\right]=0.
\label{expansion1}
\end{eqnarray}
Next, introducing the expansion $P=P_1+\epsilon P_2 +\cdots$,  
and integrating Eq.~(\ref{expansion1}) once in $T$, at order $\mathcal{O}(\epsilon^2)$ 
we obtain the following KdV equation for $P_1$:
\begin{eqnarray}
P_{1X}-\frac{\Omega^2}{2}(1-\alpha)P_{1TTT}-\alpha P_1P_{1T}=0.
\label{kdv}
\end{eqnarray}
To this end, using the soliton solution of Eq.~(\ref{kdv}) for $P_1$, 
namely $P_1=6\kappa\Omega^2{\rm sech}^2(T-X/V)$ (where $V^{-1}=2\Omega^2 \kappa \alpha$), 
we can write the approximate KdV soliton solution for $p(x,t)$ as follows:
\begin{eqnarray}
\!\!\!\!\!\!\!\!
\frac{p(x,t)}{p_0}&\approx &3\epsilon\kappa(\tilde{\omega}_0/\omega_0)^2~{\rm sech}^2
\left[\sqrt{\epsilon}\tilde{\omega}_0\left(t-\frac{x}{v}\right)\right], 
\label{kdvsoliton}
\end{eqnarray} 
where the velocity of the KdV soliton is given by 
\begin{equation}
v \approx c_0\sqrt{a} (1+2\epsilon\Omega^2 \kappa \alpha).
\label{kdvvel}
\end{equation}
It is observed that the amplitude of the normalized pressure $p/p_0$ is now of 
order $\epsilon \Omega^2$ and, thus, KdV solitons are of smaller amplitude than the 
Boussinesq-type solitons [cf. Eq.~(\ref{exactsoliton})]. In fact, 
the KdV soliton (\ref{kdvsoliton}) can be obtained as the small-amplitude limit of 
Eq.~(\ref{exactsoliton}), corresponding to $\delta=\sqrt{\epsilon} \ll 1$ (and, accordingly, 
the velocity (\ref{bousvel}) is reduced to (\ref{kdvvel}) in the same limit).

\begin{figure}[tbp]
\includegraphics[width=8cm]{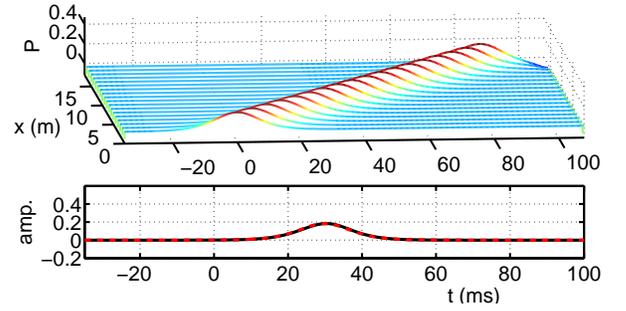}
\caption{
\label{kdvfig} 
Same as in Fig.~\ref{exact1}, but for an initial condition 
corresponding to a KdV soliton of the form of Eq.~(\ref{kdvsoliton}).}
\end{figure}

The evolution of the small-amplitude KdV soliton was also studied numerically: 
in Fig.~\ref{kdvfig} we show the result of a direct numerical simulation, 
for the same parameters as in Fig.~(\ref{exact1}), where the initial condition (at the first site 
as before) for Eq.~(\ref{dimens1}) was the KdV soliton (\ref{kdvsoliton}) at $x=0$, 
with an amplitude $\epsilon\Omega^2=0.05$. It is observed that the KdV soliton is 
much more robust, and no noticeable emission of radiation occurs; this is natural as, in this case,  
the KdV Eq.~(\ref{kdv}) is the long-wavelength far-field limit of Eq.~(\ref{dimens1}) as 
mentioned above. The analytical result for the temporal soliton profile (cf. bottom panel of the 
figure) is found to be in excellent agreement with the numerically obtained solution. 
Notice that the KdV solitons were found to be robust for propagation distances 
of the order of $60$~m (which was the distance used in the simulations).

It is interesting to compare the above approximate KdV soliton solution with 
the corresponding solution discussed in Refs.~\cite{sugi1,sugi2,sugi3}. In both cases, 
the soliton amplitude is analogous to the square root of the soliton inverse width, 
and also analogous to the geometrical factor $\kappa$. Additionally, both in our 
case and in Refs.~\cite{sugi1,sugi2,sugi3}, the KdV solitons were obtained in 
the same asymptotic limit of small amplitude and large width.

We complete this subsection by noting the following: if the initial condition for 
Eq.~(\ref{dimens1}) is fixed (i.e., the spectral width $\tilde{\omega}_0$ and amplitude are 
fixed) then the soliton amplitude will also be fixed. Nevertheless, if the cavity length $h$ 
is increased then the soliton width $w=(\delta \tilde{\omega}_0)^{-1}$ [cf. Eq.~(\ref{exactphys})] 
is also increased (this occurs for both the Boussinesq-like and KdV solitons). This theoretical 
prediction -- which is based on our analytical approximations -- is in accordance with the 
numerical and experimental results shown in Figs.~\ref{expernumer} and \ref{exper2}, respectively; 
for the latter, however, the presence of dissipation results -- additionally -- in unequal soliton amplitudes.

\subsection{NLS envelope solitons}

Our analytical approach allows us to predict still another type of soliton solutions, 
namely {\it envelope solitons} of the bright and dark type \cite{kiag}, that can be supported  
in the acoustic waveguide structure under consideration. In particular, in this Section we will 
show that such solitons can be found as approximate solutions of the nonlinear 
evolution equation~(\ref{eq_final_d}). Our methodology relies on the use of the multiple scales 
perturbation method \cite{jefkaw}, by means of which Eq.~(\ref{eq_final_d}) is reduced to an 
effective NLS equation; then, employing the latter, we identify 
parameter regimes envelope bright or dark acoustic solitons can be formed in our setting. 

We start our analysis by introducing the slow variables
\begin{eqnarray}
\chi_n=\epsilon^n\chi, \quad\tau_n=\epsilon^n\tau, \quad n=0,1,2,\ldots 
\label{slow}
\end{eqnarray}
where parameter $\epsilon$ is the one appearing in Eq.~(\ref{eq_final_d}), 
and will again be treated as a formal small parameter; furthermore, we express $P$ as an 
asymptotic series in $\epsilon$:
\begin{equation}
P=P_0+\epsilon P_1+\epsilon^2 P_2+\ldots, 
\label{asexp}
\end{equation}
where the unknown real functions $P_n$ ($n=0,1,2,\ldots$) depend on the variables (\ref{slow}). 
Then, substituting Eq.~(\ref{asexp}) into Eq.~(\ref{eq_final_d}), and using Eq.~(\ref{slow}), we 
obtain a hierarchy of equations at various orders in $\epsilon$ (see Appendix A).

In particular, at the leading order, i.e., at $\mathcal{O}(1)$, the resulting equation 
[cf. Eq.~(\ref{ord1}) in Appendix A] corresponds to the linear limit of Eq.~(\ref{eq_final_d}); 
this equation possesses plane wave solutions of the form:
\begin{eqnarray}
P_0(\tau_0,\chi_0,\tau_1,\chi_1,\ldots)&=&\Phi(\tau_1,\chi_1,\tau_2,\chi_2,\ldots) \nonumber \\
&\times& \exp\left[i\theta(\tau_0,\chi_0)\right]+{\rm c.c.},
\label{sol0}
\end{eqnarray}
where $\Phi$ is the unknown envelope function of $P_0$, the phase $\theta(\tau_0,\chi_0)$ is 
given by $\theta(\tau_0,\chi_0)=k\chi_0-\omega\tau_0$, while $k$ and $\omega$ 
satisfy the linear dispersion relation --cf. Eq.~(\ref{dispers2}). 

Next, at the order $\mathcal{O}(\epsilon)$, the solvability condition for the 
corresponding equation [cf. Eq.~(\ref{ord2}) in Appendix A] is $\tilde{L}_1 P_0=0$; 
this condition is nothing but the requirement that the secular part 
(which is in resonance with $\tilde{L}_0 P_1$) 
vanishes. This condition yields the following equation:
\begin{equation}
\left(k'\frac{\partial}{\partial \tau_1} - \frac{\partial}{\partial \chi_1}\right)
\Phi(\chi_1, \tau_1, \ldots)=0,
\label{o1}
\end{equation}
where $k'\equiv \partial k/\partial \omega$ is the inverse group velocity. Equation~(\ref{o1}) is 
satisfied as long as $\Phi$ depends on the variables $\chi_1$ and $\tau_1$ through the 
traveling-wave coordinate $\tilde{\tau}_1=\tau_1+k'{\chi_1}$ (i.e., $\Phi$ travels with the
group velocity), namely $\Phi(\chi_1,\tau_1,\ldots)=\Phi(\tilde{\tau}_1,\chi_2,\tau_2,\ldots)$.
Additionally, at the same order, we obtain the form of the field $P_1$, namely :
\begin{align}
&P_1=-\frac{4\alpha\omega^2(1-4\Omega^2\omega^2)}{D(2\omega,2k)}\Phi^2(\tilde{\tau}_1)e^{2i\theta}+Be^{i\theta}+{\rm c.c.}, 
\label{p2}
\end{align}
where $B$ is an unknown function that can in principle be found at a higher-order approximation. 

Finally, following a similar procedure as above, and using the functional forms of $\Phi$ and $P_1$, 
the non-secularity condition of the equation at the order $\mathcal{O}(\epsilon^2)$ 
[cf. Eq.~(\ref{ord3}) in Appendix A], yields a NLS equation for the envelope function $\Phi$:
\begin{eqnarray}
i\frac{\partial \Phi}{\partial \chi_2} 
-\frac{1}{2}k'' \frac{\partial^2\Phi}{\partial\tilde{\tau}_1^2}+q|\Phi|^2\Phi=0,
\label{nls}
\end{eqnarray} 
where the dispersion and nonlinearity coefficients are respectively given by: 
\begin{align}
&k'' \equiv  
\frac{\partial^2 k}{\partial\omega^2} \nonumber \\
&=\frac{1-k'^2(1-\Omega^2\omega^2)+\Omega^2(k^2-6\Omega^2\omega^2
-4\Omega^2\omega k k')}{k(1-\Omega^2\omega^2)}, 
\\
&q(\omega,k)=\frac{\alpha^2(1-\Omega^2\omega^2)(1-4\Omega^2\omega^2)}{3k\Omega^2(1-\alpha)}.
\end{align}

\begin{figure}[tbp]
\includegraphics[width=9cm]{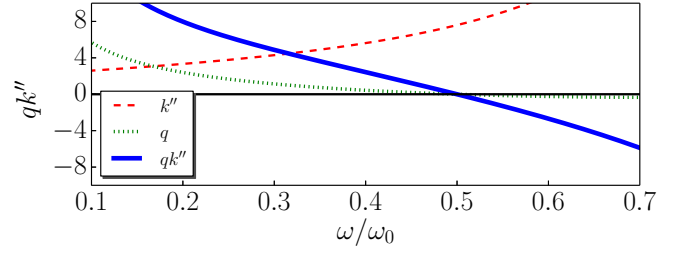}
\caption{
\label{dq} 
(Color online) The solid line shows the frequency dependence of the product $qk''$ 
of the dispersion and nonlinearity coefficients of the NLS Eq.~(\ref{nls}). 
Dashed and dotted lines shows the frequency dependence of $k''$ and $q$ respectively. 
Parameter values correspond to the experimental ones, for a Helmholtz 
resonator with cavity length $h=0.07$~m. 
}
\end{figure}

Importantly, the sign of the product $\sigma\equiv{\rm sgn}(qk'')$, 
determines the nature of the NLS equation, focusing ($\sigma=+1$) or defocusing ($\sigma=-1$) 
and, hence, the type of the soliton --bright soliton and dark soliton, respectively \cite{kiag}. 
In Fig.~\ref{dq} we show an example of the dependence of the product $qk''$ with respect to the 
frequency $\omega$, corresponding to a Helmholtz resonator of a cavity length $h=0.07$~m. 
As seen in the figure, there exist two different regimes: the low (high) frequency regime where 
$\sigma=+1$ ($\sigma=-1$) where bright (dark) solitons can be formed. 

First we consider the low frequency regime of Fig.~\ref{dq}, where the 
NLS Eq.~(\ref{nls}) is focusing and supports an exact analytical bright soliton solution of the form 
$\Phi=(\eta/\sqrt{q}){\rm sech}(\eta/\sqrt{|k''|}\tilde{\tau}_1)\exp[i(\eta^2/2)\chi_2]$.
This expression leads to an approximate bright soliton solution of Eq.~(\ref{eq_final_d}), 
which is written in terms of coordinates $\chi$ and $\tau$ as follows:
\begin{eqnarray}
P&\approx& \frac{2\eta}{\sqrt{q}}{\rm sech} \left[\frac{\epsilon\eta}{\sqrt{|k''|}}(\tau+ k'\chi)\right]
\nonumber \\
&\times& 
\cos\left[\omega\tau-\left(k-\frac{\epsilon^2 \eta^2}{2}\right)\chi\right].
\label{nlsolitonb}
\end{eqnarray}
In terms of the original space and time coordinates, the approximate envelope soliton 
solution for the pressure $p$ is the following:
\begin{eqnarray}
\frac{p(x,t)}{p_0}&\approx& \frac{2\epsilon\eta}{\sqrt{q}}{\rm sech}
\left[\frac{\epsilon\eta}{\sqrt{|k''|}}(t+ \frac{k'}{c_0\sqrt{\alpha}}\chi)\right]
\nonumber \\
&\times& \cos\left[\tilde{\omega}_0t-\left(\frac{k-(\epsilon^2 \eta^2)/2}{c_0\sqrt{\alpha}}\right)x\right]
\label{nlsolitonb}
\end{eqnarray}
where parameters $q$, $k'$, and $k''$, for a given frequency $\tilde{\omega}_0$, 
are found by using the dispersion relation in the original coordinates.

Next, we consider the high frequency regime of Fig.~\ref{dq}, where the NLS Eq.~(\ref{nls}) 
is defocusing and admits a dark soliton solution of the form 
$\Phi=\sqrt{\Phi_0} \tanh[\sqrt{\Phi_0/|k''|}\tilde{\tau}_1)\exp(-i\Phi_0 \chi_2)$.
In this case, the corresponding approximate solution of Eq.~(\ref{eq_final_d}) reads:
\begin{eqnarray}
P&\approx& 2 \sqrt{\frac{\Phi_0}{q}}\tanh\left[\sqrt{\frac{\Phi_0}{|k''|}}\epsilon(\tau+k'\chi)\right]
\nonumber \\
&\times& \cos\left[\omega\tau-(k+\epsilon^2 \Phi_0)\chi\right].
\label{nlsolitond}
\end{eqnarray}
Accordingly, the approximate dark envelope soliton solution for the pressure $p$ in the original 
coordinates is given by:
\begin{eqnarray}
\frac{p(x,t)}{p_0}&\approx& 2\epsilon\sqrt{\frac{\Phi_0}{q}}\tanh\left[\sqrt{\frac{\Phi_0}{|k''|}}\epsilon\tilde{\omega}_0\left(t+\frac{k'}{c_0\sqrt{\alpha}}x\right)\right]
\nonumber \\
&\times& \cos\left[\tilde{\omega}_0 t-\left(\frac{k+\epsilon^2 \Phi_0}{c_0\sqrt{\alpha}}\right)x \right].
\label{nlsolitond}
\end{eqnarray}
Note that both the bright and the dark solitons travel with the group velocity $1/k'$ 
(evaluated at the frequency $\tilde{\omega}_0$).

\begin{figure}[tbp]
\includegraphics[width=9cm]{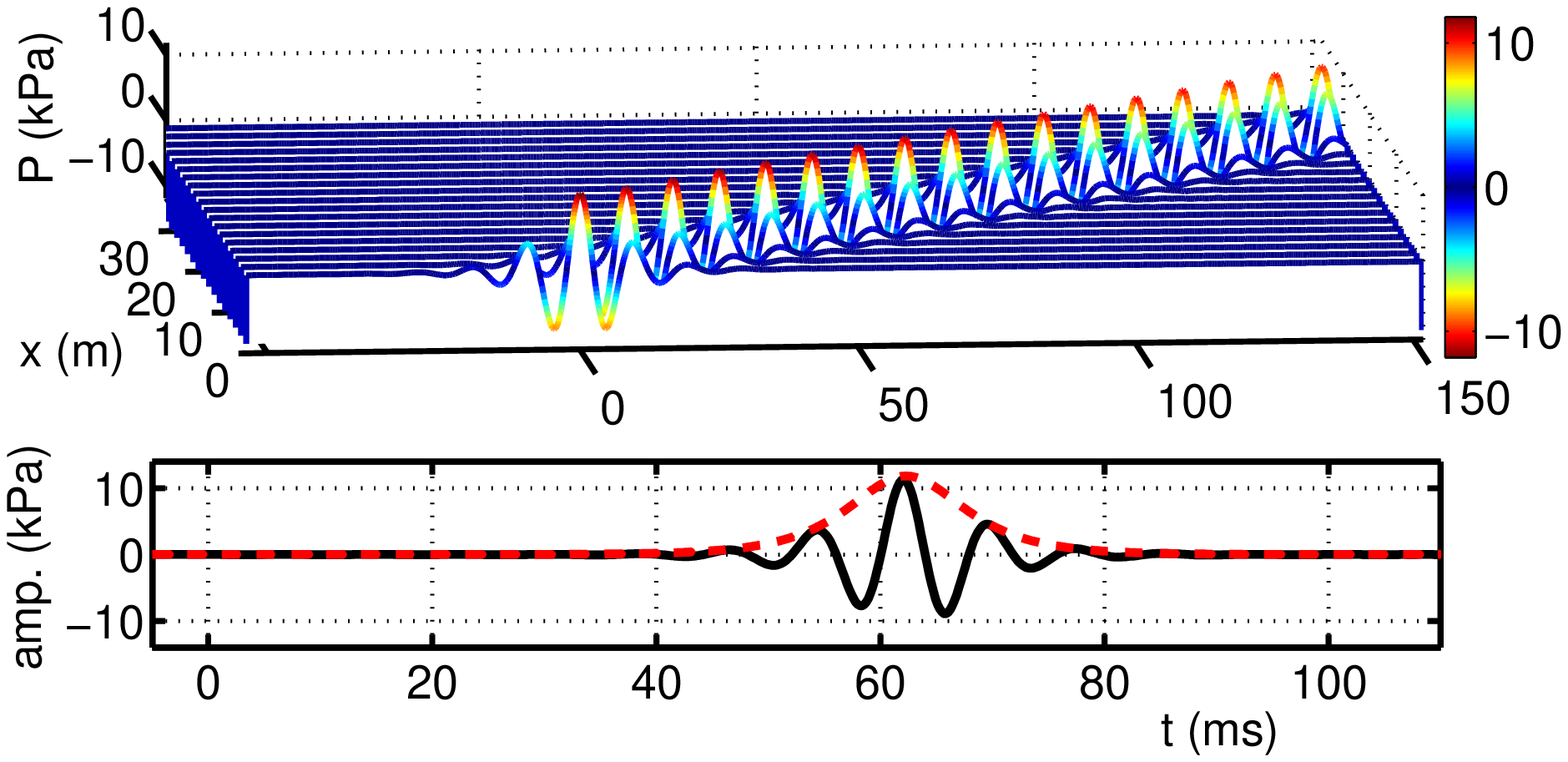}
\includegraphics[width=9cm]{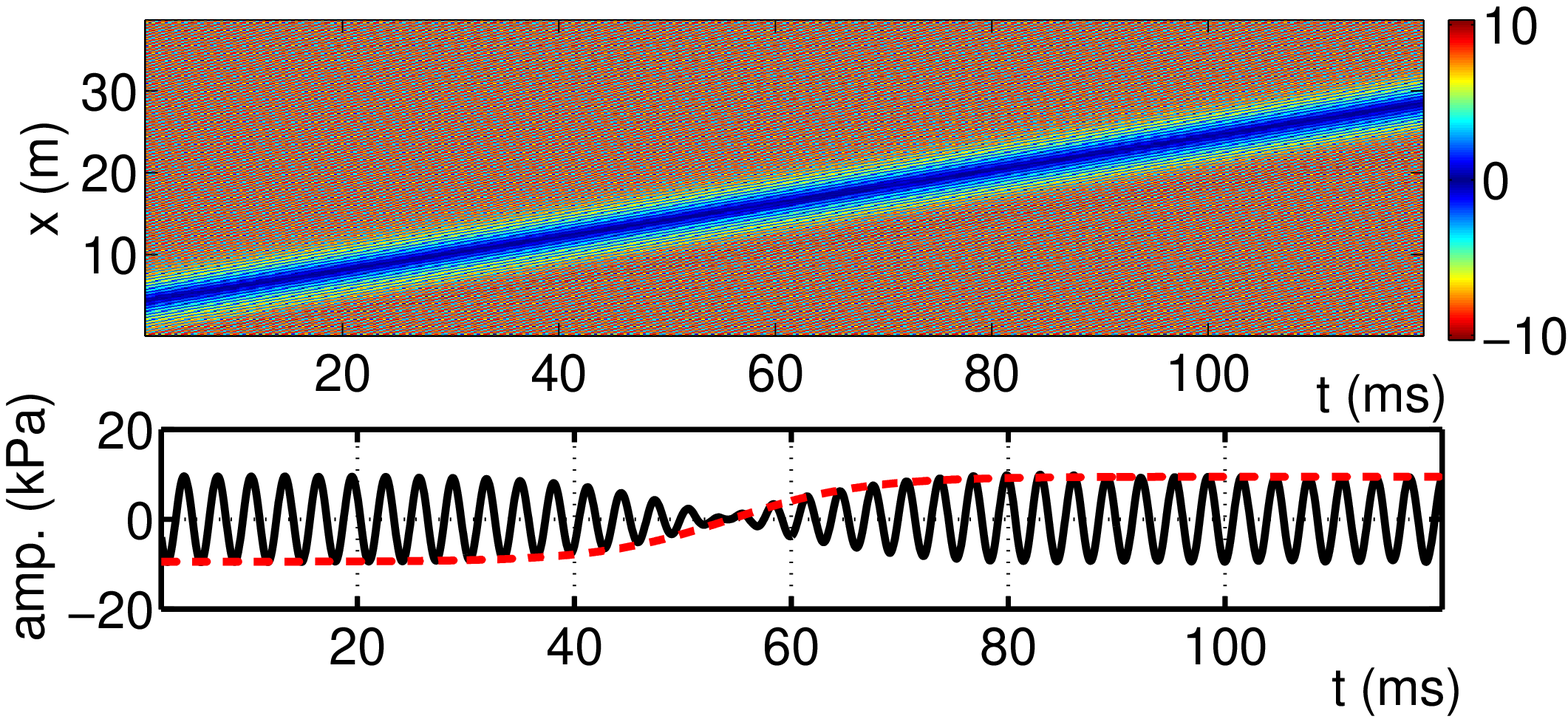}
\caption{
\label{bright} 
(Color online) 
Top panels (bottom panels): Same as in Fig.~\ref{exact1} but for 
an initial condition corresponding to a bright (dark) NLS soliton of the form of 
Eq.~(\ref{nlsolitonb}) [Eq.~(\ref{nlsolitond})]. In the second (fourth) panel, 
the red dashed line depicts the analytical result for the 
sech- (tanh-)shaped envelope of Eq.~(\ref{nlsolitonb}) [Eq.~(\ref{nlsolitond})]. 
Parameter values correspond to the experimental ones, for a Helmholtz 
resonator with cavity length $h=0.07$~m. 
}
\end{figure}

Our analytical predictions for the existence of bright and dark solitons in 
the acoustic waveguide structure at hand were also compared to direct numerical simulations. 
As in the case of the previous soliton types, we numerically integrated 
the nonlinear lattice model of Eq.~(\ref{dimens1}) using as initial conditions (at the first 
lattice site, $n=0$) the functional forms of the envelope solitons (\ref{nlsolitonb}) and 
(\ref{nlsolitond}) at $x=0$. The results are shown in Fig.~\ref{bright}, where the two 
top (bottom) panels correspond to the bright (dark) soliton, respectively. 
We have used the following parameter values: $\tilde{\omega}_0=0.2\omega_0$ and amplitude 
$\epsilon\eta=0.2$ for the bright soliton,  
$\tilde{\omega}_0=0.55\omega_0$ and $\epsilon\sqrt{\Phi_0}=0.2$ for the dark soliton. 
In the first and third panels, we show a 3D and a contour plot showing the evolution 
of these two envelope soliton types, while in the second and fourth panels we show 
the temporal profiles of the bright and dark soliton at the site $n=200$ 
(or $x=20$~m in physical units). 
It is observed that the agreement between the numerical results [solid (black) line] obtained 
in the framework of Eq.~(\ref{dimens1}) and the analytical results [dashed (red) lines depicting 
the envelopes of the two solitons] is excellent. 

\section{Conclusions and discussion}

In conclusion, we presented experimental results showing the formation of 
acoustic pulse-like solitons in an air-filled quasi-1D tube with Helmholtz resonators.  
Additionally, we proposed a transmission line (TL) approach to theoretically study our 
observations. Our model, which relied on the electro-acoustic analogy, was a 
nonlinear dynamical lattice; the latter was analyzed by both numerical and analytical techniques. 

Our numerical simulations 
produced results that were in qualitative agreement with the experimental findings. On 
the analytical side, we considered the continuum limit of the lattice model, and showed 
-- by means of dynamical systems and multiscale expansion methods -- that it can be 
reduced to celebrated soliton equations, namely a Boussinesq-type model, a KdV and a NLS 
equation. Such reductions allowed us to: 
(i) identify parameter regimes and appropriate 
spatial and temporal scales where different types of solitons can be formed, and 
(ii) derive various soliton solutions in an analytical form. 
In all cases, the analytical predictions were in excellent agreement with direct 
simulations and in qualitative agreement with the experimental observations.

In this study, our analytical approximation was simplified, due to the fact that our 
model did not take into account inherent losses in the system. This simplification, 
however, allowed us to: (a) provide analytical forms of acoustic solitons in the Helmholtz 
resonator lattice that were not available before (recall that soliton solutions 
of Refs.~\cite{sugi1,sugi2,sugi3} were presented in an implicit form, and in an explicit form 
only in some asymptotic limits for the lossless case), and (b) predict envelope solitons 
in the setting under consideration (only dark envelope solitons were 
previously predicted to occur in cylindrical acoustic waveguide structures \cite{noz}). 
Furthermore, our analytical approximation provides a clear physical picture for the 
properties of solitons in various parameter regimes and can, in principle, be used for 
other studies (thanks to the flexibility of our experimental setting) -- such as 
soliton collisions, soliton-defect interactions, 
soliton propagation in disordered lattices, and so on. 


There are many future research directions that may follow this work. 
The versatility of the experimental setting of the Helmholtz-resonators lattice, 
followed by the simplicity of the proposed nonlinear TL model, offer an attractive 
combination for a variety of future research investigations.
First, the experimental realization of envelope solitons and a systematic study of 
their properties is a particularly interesting theme. 
Also one could incorporate nonlinear elements in the
parallel branch (related to the resonators), as well as losses in the model, 
and then use asymptotic and perturbative 
techniques to capture the propagation properties of solitons, also quantitatively. Another 
interesting direction is the study of soliton formation and propagation in other 
waveguide structures, proposed or used in the context of acoustic metamaterials, with the
use of the nonlinear TL approach. In the same spirit, it would also be particularly challenging 
to extend our methodology to higher-dimensional settings. 
Pertinent studies are currently in progress and results will 
be reported in future publications.

\vspace{5pt}
\section*{Acknowledgments}
V.A. and D.J.F. acknowledge warm hospitality at LAUM, Le Mans, France, 
where most of this work was carried out. 
The work of D.J.F. was supported in part from the Special Account for 
Research Grants of the University of Athens. This study has been supported in part 
by the Agence Nationale de la Recherche through the grant ANR ProCoMedia, project ANR-10-INTB-0914.

\vspace{10pt}

\appendix
\section{Perturbation equations}

Here we present the hierarchy of equations in $\epsilon$, resulting from the substitution 
of Eq.~(\ref{asexp}) into Eq.~(\ref{eq_final_d}). More specifically, at the orders  
$\mathcal{O}(1)$, $\mathcal{O}(\epsilon)$ and $\mathcal{O}(\epsilon^2)$, we respectively obtain 
the following equatios:
\begin{eqnarray}
\!\!\!\!\!\!\!\!\!\!\!\!
&&\tilde{L}_0P_0=0,
\label{ord1} \\
\!\!\!\!\!\!\!\!\!\!\!\!
&&\tilde{L}_0P_1+\tilde{L}_1P_0=\tilde{N}_0[P_0^2],
\label{ord2} \\
\!\!\!\!\!\!\!\!\!\!\!\!
&&\tilde{L}_0 P_2+\tilde{L}_1P_1+\tilde{L}_2 P_0= \tilde{N}_0[2P_0 P_1]+\tilde{N}_1[P_0^2]. 
\label{ord3}
\end{eqnarray}
The linear operators $\tilde{L}_0$, $\tilde{L}_1$ and $\tilde{L}_2$, 
as well as the nonlinear operators $\tilde{N}_0[P]$, $\tilde{N}_1[P]$ are given by:
\begin{align}
&\tilde{L}_0=\frac{\partial^2}{\partial\tau_0^2}-\frac{\partial^2}{\partial\chi_0^2}
-\Omega^2\left(\frac{\partial^4}{\partial\tau_0^2\partial\chi_0^2}-\alpha\frac{\partial^4}{\partial\tau_0^4}\right), 
\label{L1} \\
&\tilde{L}_1=2\frac{\partial^2}{\partial\tau_0\partial\tau_1}-2\frac{\partial^2}{\partial\chi_0\partial\chi_1}
-\Omega^2\Big(2\frac{\partial^4}{\partial\tau_0^2\partial\chi_0\partial\chi_1}
\nonumber \\ 
&+
2\frac{\partial^4}{\partial\chi_0^2\partial\tau_0\partial\tau_1}-4\alpha\frac{\partial^4}{\partial\tau_0^3\partial\tau_1}\Big),
\label{L2}\\
&\tilde{L}_2  = \frac{\partial^2}{\partial\tau_1^2}-\frac{\partial^2}{\partial\chi_1^2}
+2\frac{\partial^2}{\partial\tau_0\partial\tau_2}-2\frac{\partial^2}{\partial\chi_0\partial\chi_2}
 \nonumber \\
& - \Omega^2\Bigg[\frac{\partial^4}{\partial\tau_0^2\partial\chi_1^2}
+\frac{\partial^4}{\partial\tau_1^2\partial\chi_0^2}+4
\frac{\partial^4}{\partial\tau_0\partial\chi_0\partial\tau_1\partial\chi_1}
\nonumber \\
& + \frac{2\partial^4}{\partial\tau_0^2\partial\chi_0\partial\chi_2}
+2\frac{\partial^4}{\partial\chi_0^2\partial\tau_0\partial\tau_2}\nonumber \\
&-\alpha\left( 6\frac{\partial^4}{\partial\tau_0^2\partial\tau_1^2}+4\frac{\partial^4}{\partial\tau_0^3\partial\tau_2} \right) \Bigg],
\label{L3} \\
&\tilde{N}_0[P]=\alpha\left[ \frac{\partial^2(P)}{\partial\tau_0^2}+\Omega^2\frac{\partial^4(P)}{\partial\tau_0^4} \right],
\label{N0} \\
&\tilde{N}_1[P]=\alpha\left[2\frac{\partial^2(P)}{\partial\tau_0\partial\tau_1}+4\frac{\partial^4(P)}{\partial\tau_0^3\partial\tau_1} \right].
\label{N1}
\end{align}

\end{document}